\documentclass[twocolumn,showpacs,preprintnumbers,amsmath,amssymb,prl,aps]{revtex4}
\usepackage{graphicx}
\usepackage{natbib}

\usepackage{dcolumn}
\usepackage{amsmath}

\makeatletter
\def\btt#1{\texttt{\@backslashchar#1}}
\DeclareRobustCommand\bblash{\btt{\@backslashchar}}
\makeatother

\usepackage{color}

\begin{document}

\preprint{HEP/123-qed}


\title[Short Title]{Optical Properties of Chiral Three-Dimensional Photonic Crystals\\ made from Semiconductors}

\author{Shun Takahashi$^1$}
\author{Takeyoshi Tajiri$^2$}
\author{Yasuhiko Arakawa$^3$}
\author{Satoshi Iwamoto$^{4, 5}$}
\author{Willem L. Vos$^{6}$}

\affiliation{
$^1$Kyoto Institute of Technology, Matsugasaki, Sakyo-ku, Kyoto 606-8585, Japan\\
$^2$Department of Computer and Network Engineering, The University of Electro-Communications, 1-5-1 Chofugaoka, Chofu, Tokyo 182-8585, Japan\\
$^3$Institute for Nano Quantum Information Electronics, The University of Tokyo,\\
4-6-1 Komaba, Meguro-ku, Tokyo 153-8505, Japan\\
$^4$Research Center for Advanced Science and Technology, The University of Tokyo, 4-6-1 Komaba, Meguro-ku, Tokyo 153-8505, Japan\\
$^5$Institute of Industrial Science, The University of Tokyo, 4-6-1 Komaba, Meguro-ku, Tokyo 153-8505, Japan\\
$^6$Complex Photonic Systems (COPS), MESA+ Institute for Nanotechnology, University of Twente, P.O. Box 217, 7500 AE Enschede, The Netherlands
}


\date{\today}


\begin{abstract}
We perform a theoretical and numerical study of the optical properties of both direct and inverse three-dimensional (3D) chiral woodpile structures, and a corresponding chiral Bragg stack. 
We compute transmission spectra in the helical direction for finite crystals, and photonic band structures, where we ensure the effective index of all structures to be the same. 
We find that both 3D structures show dual-band circular dichroism, where light with a particular circular polarization state - either left- or right-handed - reveals a broad stop band with major attenuation, whereas the other polarization state is transmitted nearly unimpeded. 
We observe such gaps in different frequency ranges, with alternating handedness. 
The appearance of the circular-polarized gaps is successfully interpreted with a physical model wherein the circular polarization either co-rotates or counter-rotates with the chiral structure, thereby effectively leading to a spatially-dependent or a constant refractive index, and thus to the presence or absence of a gap.
We find that the presence or absence of circularly polarized gaps is tuned by the transverse periods perpendicular to the chiral axis in both the 3D direct and inverse chiral woodpile crystal structures. 
The tunability of circular dichroism may find applications as on-chip sensors for circularly polarized light and in the on-chip detection of chiral molecules. 
\end{abstract}


\maketitle


\section{Introduction}

Inspired by the well-known occurrence of circular dichroism in chiral molecules like glucose and many amino acids [1] - where the optical response depends on the circular polarization of incident light - recent years have seen a fast-growing interest in the study and control of dichroism with chiral nanostructures in nanophotonics [2-4]. 
The favorable prospects of nanophotonics are that devices are tiny, which allows for ready on-chip integration, notably when the nanostructures are fabricated from semiconductor materials, which is the main direction of the present work. 
Simultaneously, the integration with modern microelectronics is readily foreseeable, therefore promising the advent of efficient miniature sensors of chiral molecules for myriad applications in organic and biochemistry, personal health and safety [5, 6]. 
In parallel, chiral nanostructures open the road to realizing circularly polarized laser emission without additional lossy or expensive components [7, 8], and also to polarization-controlled guiding for photonic applications [9]. 
As is relevant to topological photonics, chiral structures break spatial inversion symmetry, hence intriguing photonic Weyl points can appear in photonic band structures even in absence of an external magnetic field [10]. 
By analogy with Dirac points that allow large-volume single-mode lasers in two dimensions (2D) [11], it is proposed that Weyl points may yield similar large-volume single-mode lasers in 3D. 
Moreover, Weyl points in 3D may find applications in quantum information technology via the long-range interactions between quantum emitters [12, 13]. 
Controlling circular polarization opens intriguing prospects in light-matter physics, namely the manipulation of spin states in atomic physics and bulk semiconductors [14], or in semiconductor quantum dots and quantum wells for quantum information processing [15, 16]. 

Early work in chiral nanophotonics studied metamaterials, man-made nanostructures with building blocks that are typically much smaller than the wavelength of light. 
Following the pioneering work of Gansel \textit{et al.} broadband circular dichroism has been demonstrated with chiral metamaterials [17-23]. 
By using the effective medium approximation, the permittivity and permeability tensors are effectively determined in metamaterials [2], and the use of (noble) metals as building block serves to strongly modify these tensors, including even the off-diagonal elements [24]. 
In these structures, the light is scattered by the dielectric components [25], as well as absorbed by the metal constituents [26]. 
To circumvent absorption, attention has therefore turned to dielectric chiral photonic structures including photonic crystals, that typically have ``photonic length scales'' [27], in other words, typical feature sizes of the order of the wavelength $(a \approx \lambda)$. 

Photonic crystals are nanostructures consisting of non-absorbing dielectrics [28-31] that control light propagation by multiple Bragg interference [32-34], hence effective medium approximations are no longer valid. 
A distinguishing property of 3D photonic crystals is the fundamental control of the density of optical states (DOS), especially with the occurrence of photonic gaps that control light-matter physics such as spontaneous emission [35-41]. 
A number of studies have been reported on 3D chiral photonic crystals, notably the class of structures consisting of spirals, see Refs. [42-45]. 
In a theoretical study for dielectric contrasts representative of semiconductor nanostructures, Lee and Chan found that spiral structures reveal a complete 3D band gap or a transmission gap for either circular polarized light state separately, depending on whether the spirals are connected or spatially separated [45]. 
Another important class of 3D chiral photonic crystals consists of layers of nanorods that are stacked in the z-direction (see Fig. 1), also called spiral woodpile structures. 
A main characteristic of this class of structures is the number of layers $N$ within each spiral unit along the z-direction. 
The examples shown in Fig. 1 have $N = 3$, as will also be studied in this paper. 
With such chiral woodpile crystals broadband circular dichroism has been reported, due to circularly-polarized Bragg diffraction and apparent as circularly-polarized reflectivity gaps [46]. 

In this manuscript, we theoretically and numerically study the optical properties of chiral 3D photonic crystals made from semiconductors that reveal broad gaps. 
Open questions addressed here are whether the circularly-polarized gaps are apparent not only in direct 3D chiral woodpile crystals, but also in inverse chiral woodpile structures. 
Moreover, we investigate the role of the transverse periodicity (or even lack thereof) in the plane perpendicular to the chiral stacking direction. 
These questions are relevant for applications, since such structures are amenable to modern nanofabrication processes, in case of GaAs by micromanipulation as experimentally studied by us [31, 46-48], and in case of silicon by deep reactive ion etching through tailored etch masks [49, 50]. 
Furthermore, we study a chiral layered Bragg stack, that is the counterpart of both 3D chiral crystal structures with a homogenized lateral structure, to study the effects of the periodicity \textit{perpendicular} to the helical axis on the polarized transport. 

\section{Structures and methods}\label{sec:structures_methods}
\subsection{Chiral crystal structures}
\begin{figure}[tbp!]
\centering
\includegraphics[width=0.95\columnwidth]{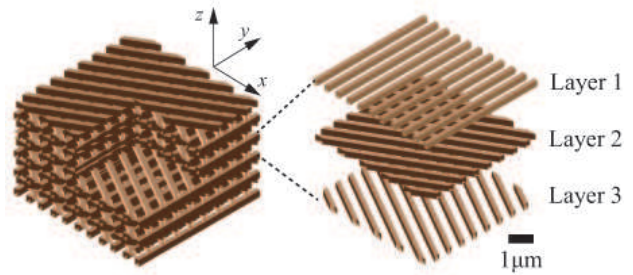}
\caption{\label{fig:chiral_woodpile}
Left: Bird's eye view of a three-dimensional (3D) three-layer left-handed chiral woodpile structure consisting of layers of nanorods that are stacked in the z-direction. 
From bottom to top ($+z$ direction), the nanorods are rotated in a counter clockwise sense, hence left handed. 
For clarity, the width of the nanorods is shown to be smaller $(w = 0.3 a)$ than in the computations. 
} 
\end{figure}

The three-layer direct woodpile chiral photonic crystal structure is schematically shown in Fig. 1 [31, 46, 48]. 
Each layer consist of an array of parallel nanorods spaced by $a$. 
The crystal structure consists of a sequence of ($N = 3$) layers whose main axes are rotated by $60^{\circ}$ compared to those of the neighboring layers. 
Thus, there is an obvious staircase-like chiral appearance, as is visible in the exploded view on the right of Fig. 1. 
As a result, the $(N + 1)$th or 4th layer is directly above the 1st layer. 
A complete helical unit is thus composed of three layers and we call its lattice spacing $c$. 

\begin{figure}[tbp!]
\centering
\includegraphics[width=0.95\columnwidth]{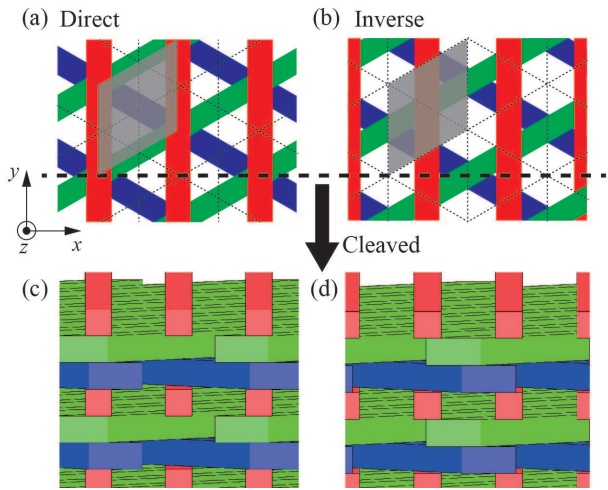}
\caption{\label{fig:direct_and_inverse}
Schematic of left-handed direct and inverse chiral woodpile structures. 
In (a,b) the structures are viewed in the $-z$ direction. 
(a) Direct woodpile structure, with the red nanorods are at the top, the green nanorods in the middle, and the blue nanorods at the bottom of the unit cell. 
Dotted lines indicate the axes of the air rods that correspond to the nanopore axes in the inverse structure. 
(b) Inverse woodpile structure, where red indicates the dielectric material in between the nanopores in the top layer, green is dielectric in the middle layer, and blue in the bottom layer. 
For clarity and transparency, the width of the dielectric is smaller $(w = 0.3 a)$ than in the calculations. 
In the inverted structure, the projection of the layers of dielectric material in one plane form a 2D Kagome lattice. 
Panels (c,d) show the (x,z)-cleaved crystal structures in a slightly tilted perspective view. 
The colors of the dielectric in the 3 layers matches the top views in (a,b). 
The cleavage plane of the front surfaces is shown in (a,b) with the horizontal dashed line. 
}
\end{figure}

The detailed structure of both the direct and inverse crystal structures is shown in Fig. 2(a-d). 
In Fig. 2(a,c) the colored units are the nanorods that are the basis of the direct structure, in Fig. 2(b,d) the colored units indicate the material in between the nanopores in the inverse structure. 
The rods and the pores are spaced by lattice spacing $a$, and we set the ratio of lattice parameters to $c/a = 1.35$, inspired by experimental work [46-48]. 
Each nanorod has a rectangular cross-section with width $w = 0.5 a$ and thickness $t = 0.45 a$. 
Similarly, each nanopore also has a rectangular cross-section with width $w = 0.5 a$ and thickness $t = 0.45 a$. 
Due to our choice $w = 0.5 a$, the inversion of high to low-index material and \textit{vice versa} keeps the average refractive index the same between both structures, which facilitates a comparison since the slope of bands away from zero frequency and the gap centers are to a first approximation the same. 
We also performed calculations for a second woodpile structure with a much larger transverse period yet the same average index, namely $a' = 2a$, $w' = 0.5a'$, $t' = t = 0.225a'$, $c' = c = 0.675a'$. 
An interesting consequence of our choice for the \textit{chiral inverse woodpile structure} is that the crossing points of consecutive pairs of dielectric material are not on top of each other anymore (projected in the z-direction) as in the direct structure; therefore, the projection of each pair of layers of dielectric material forms a two-dimensional (2D) Kagome lattice [51, 52]. 

In both 3D chiral crystal structures, the high-refractive index material is taken to be a semiconductor such as GaAs or Si with a refractive index $n = 3.4$, and the low-index material is taken to be air ($n = 1.0$).
The structural averaged index is $n_{ave} = 2.2$ within each plane of nanorods. 
Recently, the chiral direct 3D photonic crystal structure has been realized from GaAs by an advanced micro-manipulation technique [31, 34, 53-58]. 

\begin{figure}[tbp!]
\centering
\includegraphics[width=0.8\columnwidth]{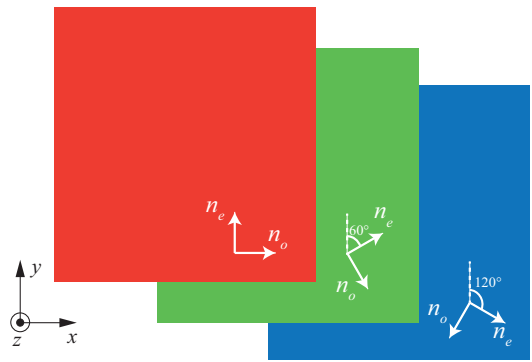}
\caption{\label{fig:bragg_stack}
Schematic of a three-layer left-handed chiral Bragg stack where the blue layer is at the bottom, the green layer in the middle and the red layer at the top. 
With increasing $z$ the extraordinary index $n_e$ rotates counter clockwise, so this Bragg stack is left handed. 
} 
\end{figure}

For comparison with the 3D chiral structures, we also study a chiral structure where the layers of nanorods or nanopores are homogenized in the (x,y) plane to yield a three-layer chiral Bragg stack, see Fig. 3 for a left-handed chiral Bragg stack. 
Each of the three constituent homogeneous layers are optically anisotropic, since the layers of both woodpile structures are also anisotropic, with ordinary index $n_o$ and extraordinary index $n_e$. 
From layer to layer with increasing $\mathrm{z}$, the optical axis is rotated by $60^{\circ}$, either clockwise or counter-clockwise depending on whether the structure is right handed or left handed, as in Fig. 3. 
The thickness of each homogeneous plate is also same as the 3D chiral woodpile structure $t = 0.45 a$ (with $c = 1.35 a$). 
In the first plate, the optical axis is along $y$. 
The anisotropic indices are chosen to obtain homogeneous counterparts of both 3D chiral photonic crystals, as follows: 
We start with one layer of nanorods that we infinitely extend in the z-direction (thus the rods are extended in the z-direction to become plates.) 
Next, we compute the 3D band structures in the z-direction, that yields two sets of linearly polarized bands without gaps (due to the structural homogeneity of the plates). 
From the slopes of these two sets of bands, we obtain the effective refractive indices of the two polarizations, that are taken as the anisotropic indices $n_{xx} = 1.6$ and $n_{yy} = 2.6$. 
The index in the z-direction is set by imposing the linear bands in the $\Gamma-Z$ direction to have the same slope as in the woodpile structures (compare Fig. 9 to Figs. 4 and 6). 
Thus the diagonal anisotropic indices are $n_{xx} = n_o$ = 1.6, $n_{yy} = n_e = 2.6$, $n_{zz} = 2.3$, the off-diagonal components are all zero ($n_{i \ne j} = 0$), and the structural averaged effective index is $n_{ave} = 2.2$. 

\subsection{Computational methods}
We calculate the photonic band structure using the well-known plane wave expansion (PWE) method [30, 59, 60] using commercially available code (Synopsis Rsoft BandSOLVE). 
We calculate the first $n = 8$ bands in the first Brillouin zone. 
For generality, in all results presented here, we normalize the frequency to the chiral lattice parameter $c$ called the reduced frequency $\tilde{\omega} \equiv c / \lambda$. 

We calculate the transmission spectra for each circularly polarized incidence by a finite-difference time domain (FDTD) method [61, 62] using commercially available code (Synopsis Rsoft FullWAVE). 
In the FDTD simulations, periodic boundary conditions are imposed in both the $x$ and $y$ directions; in the stacking direction $z$, perfectly matched layers are attached at a distance $15c$ away from the center of the structure. 
We take $M = 16$ stacked layers in the FDTD calculations, corresponding to a structure with a total thickness $L = 5 \frac{1}{3}$ unit cells, that is surrounded on both sides by air; the 16th layer has the same orientation as the 1st [63]. 
This total thickness is sufficient to detect substantial gaps in the transmission spectra. 
The incident light is a plane wave with either LCP or RCP polarization, incident normally to the top of the structure in the $-z$ direction, and detected in transmission. 
To provide frequency bandwidth, the incident light is taken to be a pulse in the time domain, with a pulse duration equal to a single optical cycle at the central frequency $\tilde{\omega} = 0.38$. 
The time-dependent electric fields of the transmitted light are recorded for both circular polarized incidences. 
We analyze the spectral responses by taking a Fourier transform of the time-dependent fields. 
The transmittance is calculated as the ratio of transmittance with and the one without a crystal structure. 

\section{Results and discussion}\label{sec:results}
\subsection{Chiral crystal direction}\label{sec:chiral}

\begin{figure}[h!]
\centering
\includegraphics[width=0.7\columnwidth]{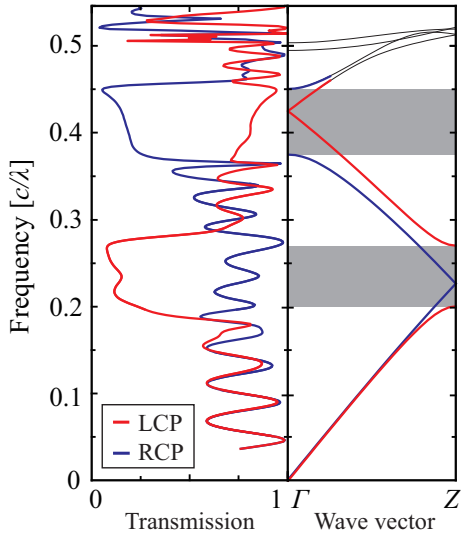}
\caption{\label{fig:chiral_woodpile_band}
(Left panel) Transmission spectrum of each circularly polarized incidence for the chiral woodpile structure composed of $N = 16$ layers. 
Circular dichroism occurs at different frequency for LCP and RCP.
(Right panel) Photonic band structure in the $z$ direction for the chiral woodpile structure. 
The lower bands are strongly polarized in LCP or RCP, which is consistent with the transmission spectra. 
}
\end{figure}

Figure 4 shows both the transmission spectrum (left panel) for circular polarized light propagating in the $z$ direction inside the 3D direct chiral woodpile structure shown in Figs. 1 and 2(a,c), as well as the band structure (right panel) in the chiral $\Gamma - Z$ high-symmetry direction. 
At low frequencies up to $\tilde{\omega} = 0.19$ the transmission shows Fabry-Perot-like oscillations, as expected for a slab-shaped sample with finite thickness. 
The transmission is the same for both left and right-handed circular polarizations, which is sensible since the periodicities are much smaller than the wavelength. 

\begin{figure}[tbp!]
\centering
\includegraphics[width=1.0\columnwidth]{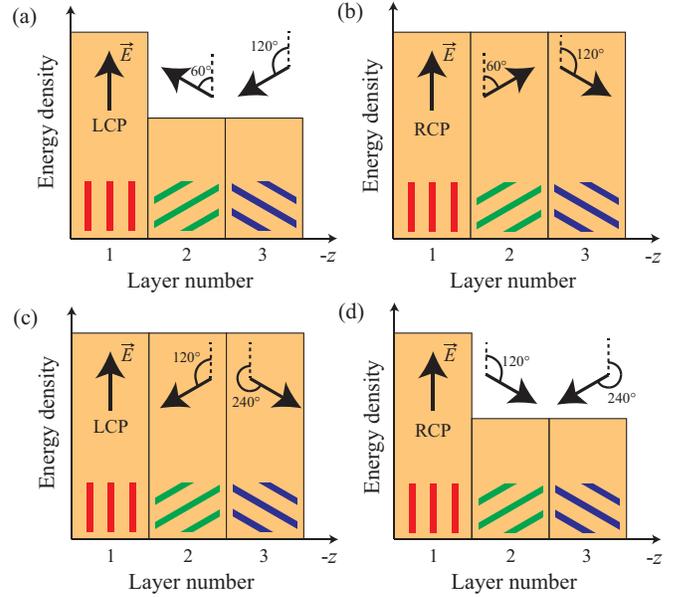}
\caption{\label{fig:energy_density}
Schematic of the energy density (sketched for the center of each layer) for (a) LCP and (b) RCP light in the 1st order Bragg condition (wavelength $\lambda = 2 n_{ave} \cdot c$). 
We illustrate a particular time (during an optical cycle) such that in the first layer, the electric field vector is parallel to the rods. 
While propagating, the electric vector of LCP (RCP) light counter-rotates (co-rotates) with the rotation of the rods. 
Thus, LCP light experiences a spatially varying refractive index, and RCP light experiences a constant refractive index during propagation. 
(c, d) Schematic energy density at the $m = 2$nd order Bragg condition ($\lambda = (2/m) n_{ave} \cdot c$) for LCP and RCP light, respectively. 
The electric vector of RCP (LCP) light counter-rotates (co-rotates) with the rotation of the rods, and experiences a spatially varying (constant) refractive index during propagation, hence a gap (no gap) appears.
}
\end{figure}

Starting at a lower edge at $\tilde{\omega}_{lo} = 0.19$ the left-hand circular polarized light has a deep and broad stopband up to an upper edge at $\tilde{\omega}_{hi} = 0.29$. 
In contrast, throughout the whole stopband, the transmission of the right-hand circular polarized light is very high. 
Therefore, in this frequency range the chiral photonic crystal shows strong circular dichroism. 
This behavior agrees very well with the band structures that show a broad stop gap for left-hand circular polarized light, whereas the bands for the right-hand circular polarized light show band folding and no stop gap at all. 


The marked presence and absence of stop gap for the two circular polarizations is understood in a real-space picture from the left-handed structure, as schematically illustrated in Fig. 5 (a) and (b). 
The figure shows the energy density for either LCP and RCP light with a wavelength $\lambda = 2 n_{ave} \cdot c$, corresponding to the first order Bragg condition [64]. 
We show the situation for a particular time during the optical cycle, such that the electric field vector is oriented in the direction parallel to the rods in the first layer. 
The electric vector of LCP (RCP) is counter-rotated (co-rotated) for the rotation of the rods. 
The LCP light experience the spatial variation of the energy density and therefore refractive index, whereas the RCP light experiences a constant energy density and thus constant refractive index $n_e$ during the propagation. 
Since the LCP light experiences a spatial variation, this simple picture leads to the occurrence of a gap, whereas the absence of spatial variation for RCP light leads to the absence of a gap, as found in Fig. 4. 


The width of the stopband corresponds to a large photonic interaction strength [65, 66] $S = \Delta \tilde{\omega} / \tilde{\omega}_{c} = 41 \%$, with $\Delta \tilde{\omega}_{s} = (\tilde{\omega}_{hi} - \tilde{\omega}_{lo})$ the band width, and $\tilde{\omega}_{s}$ the central frequency of the stopband. 
The width and center frequency of the transmission stopband are in excellent agreement with those of the stop gap in the band structure, as expected. 
The transmittance at the stop gap center is about $T = 0.12$ at $\tilde{\omega} = 0.24$, in reasonable agreement with a model based on 1D Bragg stacks, where one expects the \textit{minimum} transmission (at the stop gap center) to decrease exponentially with increasing thickness $L$ as 
\begin{align}
T(L)_{min} = \exp(-L/\ell_B), 
\label{eq:Transmission_vs_Thickness}
\end{align}
with $\ell_B$ the characteristic Bragg length that is expressed in terms of the photonic strength as 
\begin{align}
\ell_B = \frac{2d}{\pi} \frac{\omega_{s}}{\Delta \omega_{s}}, 
\label{eq:Bragglength_vs_Photstrength}
\end{align}
with $d$ the lattice spacing. 
In this model, we expect $\ell_B = 2d/(0.41 \pi) = 1.55 d$, and hence $T(L)_{min} = \mathrm{exp} (- 5.33 / 1.55) = 0.03$, which is in reasonable agreement with the results in Fig. 4 in view of the limitations of the model, namely a low index contrast, and one reciprocal lattice vector instead of a full 3D set of reciprocal lattice vectors (as illustrated in Ref. [34]). 

With increasing frequency, the transmission shows a second pronounced stopband, from $\tilde{\omega}_{lo} = 0.375$ to $\tilde{\omega}_{hi} = 0.46$, corresponding to a relative bandwidth $S = 20 \%$. 
In contrast to the first stopband, this one shows a strong attenuation for \textit{right-hand} circular polarized (RCP) light. 
This is the second order ($m = 2$) of the Bragg condition with a wavelength $\lambda = (2/m) n_{ave} \cdot c$, as schematically illustrated in Fig. 5(c, d). 
In contrast to the first order Bragg condition in (a) and (b), the electric vector of RCP (LCP) counter-rotates (co-rotates) with the rotation of the nanorods. 
The RCP light thus experiences a spatially varying refractive index $n_e (r)$, whereas the LCP light experiences a spatially constant refractive index during the propagation, leading to the appearance of the RCP stopband and no LCP stopband, in agreement with the spectral features in Fig. 4. 

The stopband shape is asymmetric, with a transmission $T = 0.18$ near the gap center ($\tilde{\omega} = 0.417$), and an absolute minimum $T = 0.05$ near the upper edge at $\tilde{\omega} = 0.45$. 
From the photonic strength, we derive $\ell_B = 2d/(0.2 \pi) = 3.18 d$, and hence predict $T(L)_{min} = \mathrm{exp} (- 5.33 / 3.18) = 0.19$, which matches well with the transmission at the gap center. 
At this time, we have no explanation for the sharply decreased transmission near the upper edge, we only note that the model invoked here does not capture such features. 

\begin{figure}[tbp!]
\centering
\includegraphics[width=0.7\columnwidth]{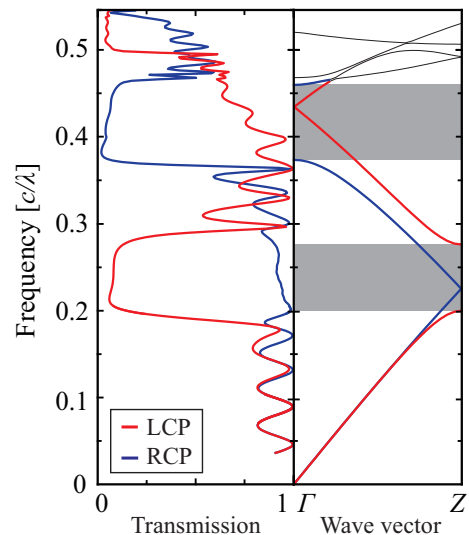}
\caption{\label{fig:kagome_band}
(Left panel) Transmission spectrum of each circularly polarized incidence for the inverse chiral woodpile structure composed of $N = 16$ layers. 
Circular dichroism occurs at different frequency for LCP and RCP.
(Right panel) Photonic band structure in the $z$ direction for the inverse chiral woodpile structure. 
The lower bands are strongly polarized in LCP or RCP, which is consistent with the transmission spectra. 
}
\end{figure}

The left panel of Fig. 6 shows transmission spectra in the $z$ direction for the \textit{inverse} chiral woodpile structure in Fig. 2. 
Because the averaged index is same as for the \textit{direct} chiral woodpile structure, the dual circular dichroism appears at the almost same frequencies. 
In brief, at lower frequencies near $\tilde{\omega} = 0.2$ to $0.28$, a broad stopband for the LCP polarization due to a 1st order Bragg condition for the LCP polarized light and simultaneously no RCP gap (see schematic Fig. 5(a,b)). 
And at higher frequencies near $\tilde{\omega} = 0.37$ to $0.46$ in Fig. 6, a broad stopband occurs for the RCP polarization due to a 2nd order Bragg condition for RCP polarized light and simultaneously no LCP gap, see Fig. 5(c,d). 
Differences occur at even higher frequencies  $\tilde{\omega}>0.5$, where the bands in the transverse directions will be seen to be different. 
This is reasonable, since small structural differences in real space correspond to slight differences at large wave vectors, which usually appear at high frequencies. 

\subsection{Transverse crystal directions}\label{sec:transverse}

The main distinction between the chiral direct and inverse woodpile structures on the one hand and the chiral Bragg stack on the other hand is of course that the woodpile structures have a 3D crystal structure, whereas the Bragg stack has only 1D periodicity in the z-direction. 
Therefore, it is expected that the (x,y) structure plays a role in the optical properties, which is the topic of this subsection.

\begin{figure}[h!]
\centering
\includegraphics[width=0.9\columnwidth]{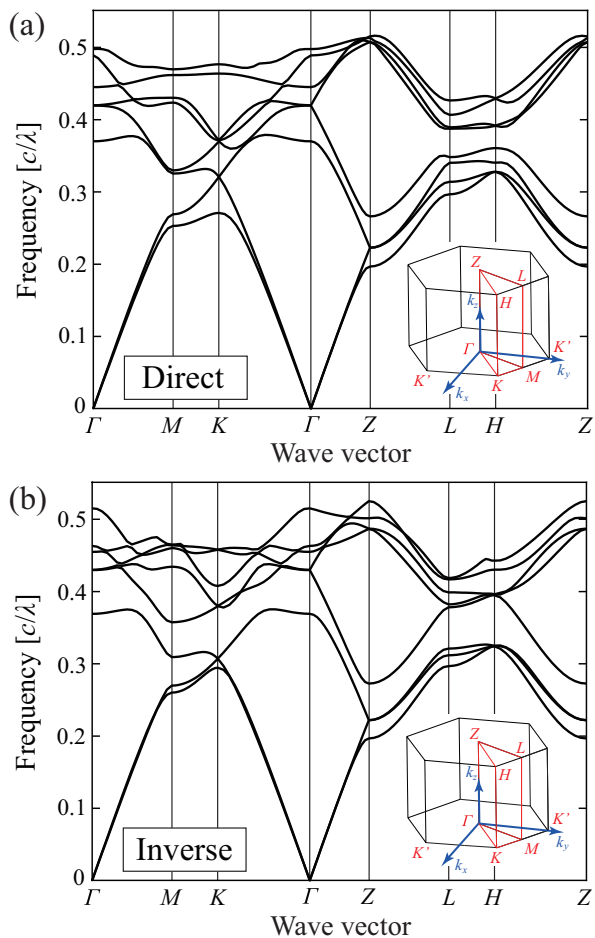}
\caption{\label{fig:bandstructure_direct_and_inverse}
3D band structures calculated for (a) the direct woodpile structure and (b) the inverse woodpile structure. 
The insets show the Brillouin zone, for simplicity only the upper half for $k_z > 0$. 
}
\end{figure}

Figure 7(a,b) show the 3D photonic band structures of the direct and inverse woodpile structures, respectively, including all directions away from the chiral $\Gamma - Z$ direction discussed above. 
Since we have chosen the width of the rods and the nanopores in the two structures to be the same, both structures have the same averaged index. 
Therefore, the two band structures strongly resemble each other, not only in the stacking direction z (correspondingly the $\Gamma - Z$ high-symmetry direction in reciprocal space), but also in the transverse directions. 
Whereas triangular lattices (relevant for the direct woodpile) and Kagome lattices (relevant for the inverse woodpile) have different photonic band structures [67], it is remarkable that both woodpile structures reveal Weyl points (at the $K$ point) [68, 69], which we attribute to the structures having the same lattice symmetry. 
The small differences of the detailed shapes of bands and the small differences for the LCP gap is attributed to the effective anisotropic index deviating in the high frequency region from the constant indices in the chiral Bragg stack. 

\begin{figure}[tbp!]
\centering
\includegraphics[width=0.7\columnwidth]{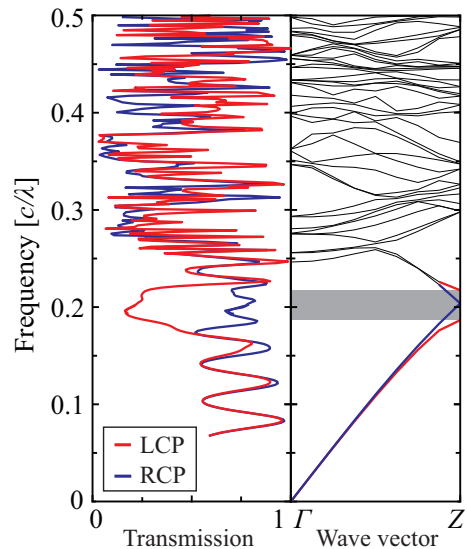}
\caption{\label{fig:woodpile_1000nm}
(Left panel) Transmission spectrum of each circularly polarized incidence for the chiral woodpile structure with a large transverse period ($a' = 2a$). 
Circular dichroism occurs only at $\tilde{\omega} = 0.2$ for RCP incidence. 
(Right panel) Photonic band structure in the $z$ direction for the inverse chiral woodpile structure. 
The lower bands are strongly polarized in LCP or RCP, which is consistent with the transmission spectra. 
} 
\end{figure}

In Fig. 8, we show the transmission spectrum and the band structures in the z-direction for a 3D chiral woodpile structure with a much larger transverse period, namely $a' = 2a$.
With increasing frequency, the transmission reveals finite-size Fabry-Perot fringes and an LCP stop band from $\tilde{\omega} = 0.17$ to about 0.23. 
The narrower polarization gap (compared to Figs. 4, 6, and 9) is explained by the smaller difference in the effective anisotropic indices $n_{xx}' = 2.18$ and $n_{yy}' = 2.79$ in the crystal with a large transverse period (as again derived from the slope of bands, see also section Structures). 
The relative difference amounts to $(n_{yy}' - n_{xx}')/(n_{yy}' + n_{xx}') = 0.12$, whereas which is twice as small 0.24 in the case of the small transverse periods $a$ discussed above. 

Simultaneously, the band structures in Fig. 8 reveal a stop gap for LCP-polarized light, but remarkably, no gap for RCP-polarized light. 
With further increasing frequency, the transmission spectrum shows strong variations: both LCP and RCP transmission quickly vary between near-0 and near-1 over narrow frequency ranges, whereby both polarizations shows different behavior. 
Simultaneously, the band structures show that many bands have appeared. 
Both features strongly differ from the ones in Fig. 4, 6, and 9), and are the result of the larger transverse lattice parameter $a$: 
with increasing $a$, the bands associated with propagation and diffraction in all other directions than the exact z-direction shift to lower frequency. 
As a result, the second order gap that is apparent in Figs. 4, 6, and 9 (around $\tilde{\omega} = 0.4$) has vanished, as it gets filled with the bands in the transverse directions. 
Consequently, this 3D crystal reveals only one circular dichroism gap (for LCP light) near $\tilde{\omega} = 0.2$. 
In both 3D crystal's band structures in Figs. 4 and 6, a few non-z diffraction bands are apparent starting at high frequencies near $\tilde{\omega} = 0.48$. 
As a further confirmation of this notion, we point out that the band structures in Fig. 9 of the 1D Bragg stack are completely "clean" all the way up to high frequencies as high as $\tilde{\omega} = 0.8$, since this 1D structure does not reveal any diffraction in non-z directions. 
Taking together all results in Figs. 4, and 6, 9, and 8, when the chiral period $c$ and the transverse anisotropic index are same, the dual circular dichroism is observed at almost the same frequency, independent on the transverse nanostructure. 

\begin{figure}[tbp!]
\centering
\includegraphics[width=0.7\columnwidth]{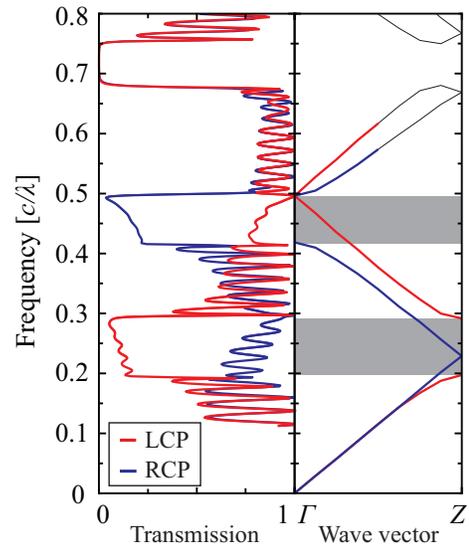}
\caption{\label{fig:bragg_stack_band}
(Left panel) Transmission spectrum of each circularly polarized incidence for the chiral Bragg stack composed of $N = 30$ layers. 
Circular dichroism occurs at different frequency for LCP and RCP.
(Right panel) Photonic band structure in the $z$ direction for the chiral Bragg stack. 
The lower bands are strongly polarized in LCP or RCP, which is consistent with the transmission spectra. 
The stop band for both circularly polarized incidence at around $\tilde{\omega} = 0.72$ is caused for $\pm 45^{\circ}$ linear polarized light, as discussed in the main text. 
} 
\end{figure}

In interpreting the role of the transverse periodicity, we also study the optical properties of a chiral structure with the simplest (homogeneous) transverse structure, namely a Bragg stack (shown in Fig. 3 with properties as similar as possible in the stacking direction as both direct and inverse woodpile structures above. 
The left panel of Fig. 9 shows transmission spectra in the stacking $z$ direction for the chiral Bragg stack. 
Near normalized frequency $\tilde{\omega} = 0.25$, RCP polarized incident light is strongly reflected by the chiral Bragg resonance, whereas LCP polarized incident light is readily transmitted though the structure. 
The converse transport - LCP being reflected and RCP readily transmitted - occurs near $\tilde{\omega} = 0.45 $.
These polarization dependencies are consistent with the band structures in the $z$ direction, shown in the right panel in Fig. 9. 
In these band structures, each band is clearly identified as RCP or LCP, as indicated by the red and blue colors, and following Lee and Chan [45]. 
The RCP polarized light has a gap near $\tilde{\omega} = 0.25$ and the LCP light near $\tilde{\omega} = 0.45$, consistent with the transmission spectra. 
This dual circular dichroism is explained by the coexistence of two helical structures in the chiral PhC, a left-handed helix with $120^{\circ}$ rotation per single layer and a right-handed helix with $60^{\circ}$ rotation per
single layer [46, 70, 71]. 

For this chiral Bragg stack, even a 3rd order stop band is apparent near $\tilde{\omega} = 0.7$, remarkably, for both LCP and RCP polarizations simultaneously. 
To understand this stopband, we consider linearly polarized light with polarization planes at $\pm 45^{\circ}$ in the $x$-$y$ plane (from which circularly polarized light can be constructed.) 
For these linear polarizations, Bragg diffraction occurs: 
While progressing in $z$ through the Bragg stack (see Fig. 3), the $+45^{\circ}$-polarized light experiences refractive indices equal to $n_a = (n_0$cos$45^{\circ}$+$n_e$cos$45^{\circ})$, $n_b = (n_0$cos$75^{\circ}$+$n_e$cos$15^{\circ})$, and $n_c = (n_0$cos$15^{\circ}$+$n_e$cos$75^{\circ})$ from the red, to the green, to the blue layer. 
Conversely, the $-45^{\circ}$-polarized light experiences refractive indices $n_a = (n_0$cos$45^{\circ}$+$n_e$cos$45^{\circ})$, $n_c = (n_0$cos$15^{\circ}$+$n_e$cos$75^{\circ})$, and $n_b = (n_0$cos$75^{\circ}$+$n_e$cos$15^{\circ})$. 
These indices are same for both polarizations but differ in their stacking order. 
Since both $\pm 45^{\circ}$-polarized light experiences the same spatial refractive index profile, it is clear that these polarizations experiences the same Bragg diffraction condition. 
Consequently, since the circularly polarizations are constructed from these orthogonal linear polarizations, both RCP and LCP light experience a same stopband near $\tilde{\omega} = 0.7$. 

\section{Summary and outlook}\label{sec:summary_outlook}
We have performed a numerical study of photonic band structures and transmission spectra in direct and inverse three-dimensional (3D) chiral photonic band gap structures. 
The structural motif of both crystal structures consists of three layers of nanorods or nanopores that are rotated by plus or minus 60 degrees in the stacking direction. 
We compared the 3D structures to chiral Bragg stacks without periodicity in the lateral, non-chiral, directions. 
Both 3D crystal structures show circular dichroism, frequency bands where light with one polarization handedness experiences a gap while the other handedness is transported nearly unimpeded. 
The frequency ranges agree with physical arguments for the spatial dielectric function that co-rotates or counter-rotates with the propagating circular polarized light. 
Moreover, both direct and inverse 3D crystal structures reveal dual-band dichroism where light with either left- or right-handed circular polarization with a stop gap at different frequency bands. 
Such polarization gaps are basically related to the helical periods with either handedness and structural averaged indices. 
Here, we find that the polarization band gaps can also be tuned by the transverse periods in the 3D direct and inverse chiral woodpile structures. 
This tunability of the circularly polarized gaps in semiconductor nanostructures can potentially be applied to design on-chip sensing devices for, \textit{e.g.}, polarization states of single photons for quantum applications, guided light for integrated photonics, and for the sensitive detection of chiral molecules for biochemistry or even point-of-care devices. 

\section{Acknowledgements}\label{sec:acknowledgements}

ST thanks Grants-in-Aid for Scientific Research, No. 21K04176. 
WLV thanks the support by NWO-TTW Perspectief program P15-36 ``Free-form scattering optics" (FFSO), the ``Descartes-Huygens" prize of the French Academy of Sciences (boosted by JMG), and the MESA$^{+}$ Institute section Applied Nanophotonics (ANP). 



\begin{thebibliography}{1}
\bibitem{footnote1}
The motivation for this choice is that when waves are propagated forward or backward through the chiral crystal, in both cases they meet a first layer of nanorods with the same orientation.
\bibitem{footnote2}
Here, $c$ is the lattice parameter in the chiral direction and should not be confused for the celerity of light.
\end{thebibliography}
\end{document}